\documentstyle[12pt,a4j]{report}
\begin{document}
\setlength{\baselineskip}{2.0pc}
\parindent = 17pt
%
\par
\begin{center}
\begin{large}
\begin{bf}
\vspace*{1.5cm}
Confined quantum systems in one dimension

and conductance oscillations in narrow channels
\end{bf}
\end{large}
\vskip 1.0cm
Karel Vacek and Ayao Okiji \par
\vskip 0.1cm
\begin{it}
Department of Applied Physics, Osaka University, Suita, Osaka 565,
Japan
\par
\end{it}
\vskip 0.1cm
Norio Kawakami \par
\vskip 0.1cm
\begin{it}
Yukawa Institute for Theoretical Physics, Kyoto University, Kyoto 606,
Japan
\par
\end{it}
\vskip 0.3cm
(Received \hskip5cm )
\end{center}
\vskip 0.5cm
\par
%
\centerline {\bf ABSTRACT } \par
\vskip 0.1cm
An exactly solvable electron model of a confined system with
inverse-square interaction is presented.
The ground state is given  by
the Jastrow-product wavefunction of  power-law form.
We discuss the results in connection with conductance oscillations
observed in semiconductor nanostructures, for which
single-electron charging effects play a crucial role.
Due to the internal spin degrees of freedom,
there appear two independent periods of the conductance
oscillations in very narrow channels even at zero temperature.
%
\newpage
\begin{center}
I. INTRODUCTION
\end{center}

One-dimensional quantum models with long-range
$1/x^2$ interaction have attracted renewed interest recently.
There are several classes of Hamiltonians which
are known to be integrable [1-4]. For instance, a family of models
with periodic boundary conditions
has been studied extensively, which gives
the models with $1/\sin ^2{x}$ interaction [2-4].  Another
interesting class of Hamiltonians has been provided by
the quantum systems confined with harmonic potential [1,2].
The effect of the internal spin-degrees of freedom
has been studied quite well for the former case:
the exact wavefunctions and the corresponding eigenenergies have been
clarified almost completely for various kinds
of multicomponent models [5-9]. However, this is not the case
for the confined systems, for which only the single-component model
has been actively investigated [1,10].

Besides intrinsically theoretical interest in such confined
electron systems with the spin degrees
of freedom, there is also a strong experimental
motivation for their investigation. For example,
experiments on one-dimensional electron systems
have been reported for very narrow channels
in Si or GaAs semiconductor nanostructures [11-14].
The conductance of a narrow channel displays periodic oscillations,
which have been ascribed to the single-electron charging
effect of the confined segment in the one-dimensional
electron gas [15-17]. A solvable electron model
with $1/x^2$ interaction  [18,19]
provides an analytical tool to investigate the
correlation effects on such experiments.

In this paper, we would like to study an exactly solvable
electron model for a {\it confined system} with the
$1/x^2$ interaction including the {\it exchange effects}.
In Section II, we introduce the model by generalizing
the original one [1] to the case including the spin
degrees of freedom. We then find the ground-state
eigenfunction in the form of the Jastrow product,
and  obtain the ground-state energy exactly.
The solution is obtained by generalizing  techniques
developed in Ref. [1-3,6].
In Section III, we apply the results to
experiments for the narrow channels in semiconductor
nanostructures, by introducing a weak coupling between the
confined system and the reservoirs.
We mainly discuss the exchange effect
on conductance oscillations, which has not
been taken into account in the previous studies
[18,19]. In particular,
we find two independent periods for the
oscillations even at zero temperature,  due to the internal
spin degrees of freedom.
Conclusion is given in Section~IV.

\vskip 8mm
\begin{center}
II.  CONFINED MODEL WITH $1/x^2$ INTERACTION
\end{center}

\begin{center}
A. Hamiltonian and Jastrow wavefunction
\end{center}

Let us consider a one-dimensional electron system confined by
the harmonic potential $\frac{1}{2}m \omega_0^2 x^2$ where
$m$ is the mass of a particle.
We introduce  the following  model Hamiltonian in units
of $\hbar^2/m$ for the system of $N$ electrons with the
inverse-square pair interaction
$$H = - \frac{1}{2} \sum_{i=1}^{N} \frac{\partial^2}{\partial x_i^2}
    + \frac{m^2 \omega_0^2}{2 \hbar^2} \sum_{i=1}^{N} x_i^2
    + \sum_{j>i} \frac{\lambda(\lambda+P_{ij}^\sigma)}{(x_j-x_i)^2}
,
\eqno(1)$$
where $\lambda \geq 0$ is the dimensionless interaction parameter and
$P_{ij}^\sigma$ is the spin-exchange operator of two particles.
The spin-exchange interaction is chosen here so as to reproduce
the non-interacting limit correctly.
Note that this interaction can lift a pathological degeneracy
of the ground state for the  multicomponent model [19]
arising from peculiar short-range
properties of $1/x^2$ interaction in one-dimension.
This point is crucial to confront the results with
conductance-oscillation phenomena, which will be mentioned
later again. The above Hamiltonian with the harmonic potential
is a generalization of the Calogero model [1], and may be
regarded as  a variant of the
SU($N$) Sutherland model introduced by Ha-Haldane for
the periodic case [6]. We will see in the next section that the model
exhibits  quite different behaviors from those
discussed in [19].

Based on the observation of  Ha-Haldane's ansatz used
for systems with periodic boundary conditions [6],
we now write down the Jastrow-type trial function
for systems with the confinement
$${\mit \Psi}(x_1 \sigma_1, \ldots ,x_N \sigma_N) =
\left \{ \prod_{j>i} |x_j-x_i|^\lambda
(x_j-x_i)^{\delta_{\sigma_j \sigma_i}}
{\rm exp}
\left [
i\frac{\pi}{2}{\rm sgn}(\sigma_j-\sigma_i)
\right ]
\right \}
\prod_{i=1}^{N} {\rm exp}(-\frac{m \omega_0}{2\hbar}x_i^2)
,
\eqno(2)$$
where ${\sigma_i}$ is the particle spin.
This wavefunction  satisfies the Pauli principle,
$M_{ij} {\mit \Psi} = - {\mit \Psi}$,
where $M_{ij}$ is a particle-exchange operator.
It is also instructive to rewrite the wavefunction out of
two parts,
$${\mit \Psi}(x_1 \sigma_1, \ldots ,x_N \sigma_N) =
\prod_{j>i} |x_j-x_i|^\lambda
{{\mit \Psi}_0} (x_1 \sigma_1, \ldots ,x_N \sigma_N),
\eqno(3)$$
where ${{\mit \Psi}_0} $ is the exact ground-state
wavefunction of the noninteracting case ($\lambda=0$).
For the above trial wavefunction, therefore, the
effect of interaction is assumed to be taken into
account completely by the Jastrow
factor $|x_j-x_i|^\lambda $, as is the
case for all other integrable $1/x^2$ systems.
In particular, the expression (3) is quite general for
the $1/x^2$ models, which is also applicable to the
periodic models by choosing an appropriate
wavefunction ${{\mit \Psi}_0}$ [6].

\vskip 8mm
\begin{center}
B. Ground-state energy
\end{center}

We now wish to show that the wavefunction (2) satisfies
the Schr\"{o}dinger equation $H {\mit \Psi} = E {\mit \Psi}$ with
the eigenenergy $E$. We limit ourselves to the proof in the sector
$x_1 \leq x_2 \leq \ldots \leq x_N$
which can be easily extended to the whole configuration space.
We start by applying the kinetic term and the confining-potential
term on the ansatz eigenfunction.
Following a technique outlined in [1], this
action turns out to be
$$\frac{1}{{\mit \Psi}}
\left [ \frac{1}{2} \sum_{i=1}^{N}
\left (-\frac{\partial^2}{\partial x_i^2}
+ \frac{m^2 \omega_0^2}{\hbar^2} x_i^2
\right )\right ]
{\mit \Psi}
= \frac{m \omega_0}{2 \hbar} \left [\lambda N ( N - 1 )
+ N_\uparrow^2 + N_\downarrow^2 \right ]- A ,
\eqno(4)$$
where the number of particles with the spin up (down) is
$N_\uparrow$
($N_\downarrow$),
$N_\uparrow + N_\downarrow = N$.
The term $A$ can be brought to the shape
$$ A =
A^{\rm (I)} + A^{\rm (II)} + A^{\rm (III)} =
\sum_{k<\ell} \frac{\lambda(\lambda-1)}{(x_k-x_\ell)^2} +
\sum_{k<\ell} \frac{2 \lambda \delta_{\sigma_k \sigma_\ell}}
{(x_k-x_\ell)^2} +
\sum_{i \neq k \neq \ell}
\frac{\lambda \delta_{\sigma_i \sigma_k}}{(x_i-x_k)(x_i-x_\ell)}
{}.
\eqno(5)
$$
Consider next the interaction term.
In order to determine the action of the spin-exchange operator
on the ansatz eigenfunction,
we first calculate  the action of the coordinate-exchange
operator $Y_{k\ell}$,
$$\frac{Y_{k\ell}{\mit \Psi}}{{\mit \Psi}} =
(-1)^{\delta_{{\sigma_k}{\sigma_\ell}}}
\prod_{i \neq k\ell}
\left ( \frac{x_i-x_\ell}{x_i-x_k} \right )
^{\delta_{{\sigma_i}{\sigma_k}}-\delta_{{\sigma_i}{\sigma_\ell}}} .
\eqno(6)
$$
According to the antisymmetric nature of
electrons, we rewrite the action of the spin-exchange operator
as $P_{k\ell}^\sigma {\mit \Psi}
= Y_{k\ell} M_{\ell k} {\mit \Psi}
= - Y_{k\ell} {\mit \Psi}
$ for  any $k<\ell$.  Hence the interaction
term on the ansatz eigenfunction gives
$$B = \frac{1}{{\mit \Psi}}
\left[ \sum_{j>i} \frac{\lambda(\lambda+P_{ij}^\sigma)}{(x_j-x_i)^2}
\right ] {\mit \Psi}
= B^{\rm (I)} + B^{\rm (II)} + B^{\rm (III)}
$$
$$ =
\sum_{k<\ell} \frac{\lambda(\lambda-1)}{(x_k-x_\ell)^2} +
\sum_{k<\ell} \frac{2 \lambda \delta_{\sigma_k \sigma_\ell}}
{(x_k-x_\ell)^2}+
\sum_{k<\ell} \frac{\lambda}{(x_k-x_\ell)^2}
\left [1-\prod_{i \neq k\ell}
\left (\frac{x_i-x_\ell}{x_i-x_k} \right )
^{\delta_{{\sigma_i}{\sigma_k}}-\delta_{{\sigma_i}{\sigma_\ell}}}
\right ]
\left (1 - \delta_{\sigma_k \sigma_\ell} \right ).
\eqno(7)$$
One can see from Eqs. (4), (5) and (7) that the application
of the Hamiltonian produces not only the constant term
but also the multiparticle terms.
If these multiparticle terms cancel each other completely
($A-B=0$), we get the eigenenergy by the first term of (4).
We will show that this is indeed the case.

We would like to prove here that $A-B=0$  by
employing a technique analogous to
that used for the periodic case [6].
Note first that the spin-independent parts
(namely the first terms $A^{\rm (I)}$ and $B^{\rm (I)}$)
and the parts for the same spins $\sigma_k=\sigma_\ell$
(namely the second terms $A^{\rm (II)}$ and $B^{\rm (II)}$)
of the expressions for $A$ and $B$ are the same.
Furthermore, by using the following relation,
$$ \left (
\frac{x_i-x_\ell}{x_i-x_k}
\right )
^{\delta_{{\sigma_i}{\sigma_k}}-\delta_{{\sigma_i}{\sigma_\ell}}}
= 1+ (x_k - x_\ell)
\left ( \frac{\delta_{\sigma_i \sigma_k}}{x_i - x_k}
-
\frac{\delta_{\sigma_i \sigma_\ell}}{x_i - x_\ell}
\right ) ,
\eqno(8) $$
where $\sigma_k \neq \sigma_\ell$, we can rewrite
the part containing pairs with different spins,
$$
B^{\rm (III)} =
\frac{1}{2}
\sum_{k \neq \ell} \frac{\lambda}{(x_k-x_\ell)^2}
\left \{ 1 - \prod_{i \neq k\ell}
\left [ 1+ (x_k - x_\ell)
\left ( \frac{\delta_{\sigma_i \sigma_k}}{x_i - x_k}
-
\frac{\delta_{\sigma_i \sigma_\ell}}{x_i - x_\ell}
\right ) \right ] \right \}
\left (1 - \delta_{\sigma_k \sigma_\ell} \right ).
\eqno(9) $$
So, our objective is now to rearrange the expression
for $B^{\rm (III)}$ in order to cancel it
with the  $A^{\rm (III)}$ term.
To this end let us expand the above product over $i \neq k \ell$
in expression (9).
One finds that the constant term $B^{\rm (III)}_0$ vanishes,
and the linear term $B^{\rm (III)}_1$  is equal to $A^{\rm (III)}$.
The terms of higher order, $B^{\rm (III)}_q$, $ 2 \leq q \leq N-2 $,
can be simplified by introducing {\it clusters} of $q+2$ particles
where $r+1$ particles have the spin up and $s+1$ particles
have the spin down, $r+s=q$.  Namely, let us write
$B^{\rm (III)}_q$, $q \geq 2$, as a sum over such clusters
$$
B^{\rm (III)}_q = -
\frac{\lambda}{2}
\frac{1}{q!}
\sum_{k \neq \alpha_1 \neq \ldots \neq \alpha_r \neq
    \ell \neq \beta_1 \neq \ldots \neq \beta_s }
b_q (k \alpha_1 \ldots \alpha_r \ell \beta_1 \ldots \beta_s)
,
\eqno(10)
$$
where
$$
b_q (k \alpha_1 \ldots \alpha_r \ell \beta_1 \ldots \beta_s) =
(-1)^s
(x_k-x_\ell)^{r+s-2}
\prod_{i=1}^r \frac{1}{x_{\alpha_i}-x_k}
\prod_{j=1}^s \frac{1}{x_{\beta_j}-x_\ell}
\eqno(11)
$$
for $\sigma_k = \sigma_{\alpha_1} = \ldots = \sigma_{\alpha_r} \neq
\sigma_\ell = \sigma_{\beta_1} = \ldots = \sigma_{\beta_s} $.
As shown in Appendix, the sum of $b_q$ ($q \geq 2$) over
the permutations $k \leftrightarrow \alpha_i$, $i=1,2,\ldots,r$, and
$\ell \leftrightarrow \beta_j$, $j=1,2,\ldots,s$,
is zero within the particles of the same spin in the
cluster of $q+2$ particles:
$$ \sum_{\{k \alpha_1 \ldots \alpha_r \} }
\sum_{\{\ell \beta_1 \ldots \beta_s \} }
b_q (k \alpha_1 \ldots \alpha_r \ell \beta_1 \ldots \beta_s) = 0
,
\eqno(12)$$
where $q \geq 2$ is fixed, and $r,s$ are arbitrary
(satisfying only $r+s=q$). Consequently, we have shown that
contribution from the terms of higher order
$B^{\rm (III)}_q$, $2 \leq q \leq N-2$, is identically zero, and
that $A-B=0$ holds.
Therefore, we find the wavefunction (2) to be the
eigenfunction of the Hamiltonian (1).

The eigenenergy $E (N_\uparrow , N_\downarrow)$
is expressed,  according to expression (4),
by a remarkably simple formula
$$
E (N_\uparrow , N_\downarrow)
=
\frac{1}{2} \hbar \omega_0
\left [
\lambda
 (N_\uparrow + N_\downarrow)
 (N_\uparrow + N_\downarrow - 1)
+ N_\uparrow^2 + N_\downarrow^2
\right ]
{}.
\eqno(13)
$$
This expression also ensures the result predicted in [20].
Although it is not easy to prove rigorously that the above eigenstate
indeed corresponds to the {\it ground state}, one can see evidence for it
in several limiting cases.
In the limit of $\omega_0 \rightarrow 0$,
the Hamiltonian (1) becomes a positive definite operator with the lowest
eigenvalue of zero, as was shown by Polychronakos [8].
One can see that the formula (13) actually produces the value of zero
in such limiting case, which proves that the energy (13) indeed corresponds
to the ground-state energy.
Also, in case of electrons with a single spin component, it is seen
that the wavefunction (2) with the energy (13) reduces to the exact
ground-state wavefunction [1,2,8].
{}From these observations we believe that the eigenfunction (2) with the
energy (13) generally describes the exact ground state of the
Hamiltonian (1).

The ground-state eigenenergy $E_0(N_\uparrow,N_\downarrow)$
is characterized by
$N_\uparrow=N_\downarrow$ (for $N$ even) or
$N_\uparrow=N_\downarrow \pm 1$ (for $N$ odd).
The ground state is also the state with the minimal spin.
In the limit of weak interactions ($\lambda \rightarrow 0$),
the formula (13) reproduces the eigenenergy expected
for non-interacting electrons with the internal
spin degrees of freedom,
$E(N_\uparrow,N_\downarrow, \lambda \rightarrow 0)
=\frac{1}{2} \hbar \omega_0 (N_\uparrow^2+N_\downarrow^2)$.
It is instructive to note here that the correct
non-interacting energy cannot be reproduced
if we omit the exchange term from the Hamiltonian, as has been
done in [19].
This pathological behavior comes from a peculiar property of $1/x^2$
interaction (short-range divergence property) in one dimension,
which gives rise to a large number of degeneracies of the ground state.
Therefore,
the exchange term used in this paper is crucial to correctly
describe interacting electrons in terms of $1/x^2$ interaction
from the weakly to the strongly correlated regimes [6].

\vskip 8mm
\begin{center}
C. Relationship to the periodic model
\end{center}

Before moving to discussions of conductance oscillations,
let us glance at the relationship to the model
with periodic boundary conditions [6].
When the number of electrons is increased by
$n_\uparrow + n_\downarrow$ from
$N_\uparrow=N_\downarrow=N_0$,  the energy increment
$\epsilon$ quadratically proportional to $n_\uparrow $
and $n_\downarrow$ is written in a simple form
(linear part gives the chemical potential) [20],
$$
\epsilon (\vec n)= {1 \over 2} \hbar \omega_0 \,
\vec n^t {\bf A} \vec n,
\eqno{(14)} $$
with the $2 \times 2$ matrix
$$ {\bf A}=
\left( \matrix {\lambda+1  & \lambda  \cr
     \lambda   &  \lambda+1 \cr}
      \right).
\eqno{(15)}
$$
On the other hand, the low-energy excitation spectrum for
the periodic $1/x^2$ ring of
circumference $L$ is classified in terms of the velocity $v$ of
elementary excitations [6],
$$ \epsilon= {2 \pi v \over L} \, [{1 \over 4}
\vec n^t {\bf A} \vec n
+   \vec d^t {\bf A}^{-1} \vec d],
\eqno{(16)}
$$
where the two-component vector $\vec d$
consists of the quantum numbers $d_\sigma$ ($\sigma= \uparrow,
\downarrow$) which carry the momentum change $2k_F d_\sigma$
with the Fermi energy $k_F$.
It is remarkable that the excitation spectrum
is classified solely by the matrix ${\bf A}$,
which is consistent with requirement of the conformal invariance
for U(1) gaussian models [21]. In that case, the interaction parameter
$\lambda$ features the $c=1$ conformal critical line.
Furthermore by imposing the chiral
constraint to the periodic model,  consider only
right-going (or left-going) electrons.
The excitation spectrum (16) is then modified into
$$
\epsilon= { \pi v \over L} \, \vec n^t {\bf A} \vec n,
\eqno{(17)}
$$
which has the same form as (14).
It can be said, therefore, that the present model with
confining potential exhibits the low-energy spectrum
essentially same as for the chiral version of the
periodic model, when we take the limit of
$\omega_0\rightarrow 0$ and $L \rightarrow \infty$
with keeping $\hbar \omega_0 =2 \pi v /L$. We note that
the critical behavior of the energy spectrum (17) is described by
holomorphic piece of U(1) Kac-Moody algebra (chiral Luttinger liquid).

\vskip 8mm
\begin{center}
III. CONDUCTANCE OSCILLATIONS IN NARROW CHANNELS
\end{center}

In this section,
we will apply our model to the transport of electrons
through a narrow channel of the semiconductor nanostructure.
The one-dimensional electron gas is confined
by impurities [12] or by constrictions [13]
to a finite segment of the channel.
We would like to model correlated electrons in such segment
by Hamiltonian (1), where we substitute the mass $m$ of a particle
by the effective mass $m^*$ of electrons in Si or GaAs [14].
Experimentally, both the confining potential and
the electron-electron interaction potential are difficult to estimate.
Generally, these potentials are believed to be device dependent.
According to this uncertainty in the shape of the confining and
the interaction potentials, the parameters $\omega_0$ and
$\lambda$ may be treated empirically
to achieve the best fit to the experiment.
We wish to mention here that applications of the $1/x^2$ models
to conductance oscillations have been done in [18,19],
and several characteristic properties have been explained.
However, the models have been restricted to
spinless fermions so far, in which the exchange effect
is completely neglected.

We now introduce a weak coupling via the confining
constrictions or impurities between the segment and
two reservoirs [22] of the adjustable chemical potential
$\mu_1 \approx \mu_2$, $\mu=(\mu_1+\mu_2)/2$.
Let the segment initially contain $N$ electrons.
The $(N+1)$th electron can enter the segment via a resonant tunneling.
The resonance occurs whenever the average chemical potential $\mu$ of
reservoirs aligns with the lowest unoccupied level of the segment.
The resonance appears as a peak in the conductance.
The position of a peak for $N \rightarrow N+1$
corresponds to a chemical potential
$\mu(N)=E_0(N+1)-E_0(N)$. The zero-temperature spacing $\delta$
of two successive peaks for
$N\rightarrow N+1$ and $N+1\rightarrow N+2$
in the conductance oscillations is given by
$\delta(N)=\mu(N+1)-\mu(N)$.  It should be noted that
in contrast to previous models [18,19],
$\delta$ can take  several different values
because of the spin-dependent exchange effects.

Let us now suppose  that the
system is {\it initially} in the ground state
with the even number of electrons
($N_\uparrow=N_\downarrow=\frac{1}{2}N$).
The increase of the chemical potential $\mu$ leads to the
first conductance peak at the position
$\mu(N_\uparrow+1, N_\downarrow)$.
The further increase of the chemical potential $\mu$ causes
the second peak at $\mu(N_\uparrow+1, N_\downarrow+1)$.
The spacing between the first and the second peaks is then
computed as $\mu(N_\uparrow+1, N_\downarrow+1)-
\mu(N_\uparrow+1, N_\downarrow)=\hbar \omega_0 \lambda$.
Similarly, the spacing between the second and the third peaks is
$\mu(N_\uparrow+2, N_\downarrow+1)-
\mu(N_\uparrow+1, N_\downarrow+1)=\hbar \omega_0 (\lambda+1)$,
the spacing between the third and the fourth peaks is
$\mu(N_\uparrow+2, N_\downarrow+2)-
\mu(N_\uparrow+2, N_\downarrow+1)=\hbar \omega_0 \lambda$,
etc.
It is seen that there appear {\it two independent periods} of
the conductance oscillations,
$$
\delta_1 = \hbar \omega_0 \lambda, \hskip 5mm
\delta_2 = \hbar \omega_0 (\lambda+1),
\eqno{(22)}
$$
reflecting {\it the internal spin degrees of freedom}.
These two periods of conductance oscillations appear alternatively;
two peaks with spacing $\delta_1$ are surrounded from each side
by a peak at spacing $\delta_2$ and vice versa.
This is a fundamental extension beyond the $1/x^2$ models discussed
by Tewari [18] and also beyond the Johnson-Payne model [19]
which is essentially the same as the spinless fermion model
with a single period of the conductance oscillations
$\delta = \hbar \omega_0 (\lambda+1)$.

For the empirical fit of parameters used by Johnson and Payne
$\delta \approx 8.5\hbar \omega_0$ [19],
we obtain two periods
$\delta_1=7.5\hbar\omega_0$ and
$\delta_2=8.5\hbar\omega_0$,
implying the correction due to the exchange effect is rather small
for these parameters.
As the interaction becomes weaker (smaller $\lambda$), however,
the above exchange effect becomes conspicuous,
making two periods more distinct.
When the interaction strength
takes the vanishing value ($\lambda \rightarrow 0$),
the present model reproduces
the results of free electrons including the spin degrees of freedom
in the harmonic potential
($\delta_1 \rightarrow 0$, $\delta_2 \rightarrow \hbar \omega_0$).

For finite temperatures, our solution to the model Hamiltonian
must be extended to include the excited states.
We expect that at finite temperatures
the peaks become broadened [18],
their height can be non-trivially changed [17],
and their spacing may be also changed
due to the presence of excited levels [19].
The number of independent periods may increase from
two at zero temperature
to several periods at finite temperatures.

Finally,
we would like to mention briefly relationship of our results to
experiments on periodic conductance oscillations
in narrow channels [11-14].
Experimentally,
there are reports that the period of conductance resonances is
not only a single one [11,14].
In addition to the dominant periodicity in the Fourier spectrum,
several peaks have been often seen at positions corresponding
to frequencies which are not harmonics of the dominant one.
The origin of multiple periods, which have been sometimes observed
experimentally,
has been explained by the possible existence of several segments
with comparable pinning energies [11].
It can be interesting to check experimentally whether one may observe
two periods in the conductance oscillations
due to the exchange effect at very low temperatures.
If so, it may provide further knowledge
of the exchange-correlation effects in one-dimensional systems.

\vskip 8mm
\begin{center}
IV. CONCLUSION
\end{center}

We have studied  a solvable electron  model for a confined
one-dimensional system interacting by inverse-square interaction.
We have obtained the exact ground-state energy with the
Jastrow-product eigenfunction.
The results have been applied to the analysis of
the periodic conductance oscillations in narrow channels.
Though the present model includes the interaction
and the exchange effect in a specific way,
it clearly demonstrates that two kinds of oscillations can be expected
in the single-electron charging phenomena due
to the internal spin degrees of freedom.

\vskip 8mm
\begin{center}
ACKNOWLEDGMENTS
\end{center}

We gratefully acknowledge fruitful discussions with
H. Kasai, Y. Kuramoto,
H. Nakanishi, S. Suga, and S.-K. Yang.
This work is partly supported by
Grant-in-Aid from the Ministry of Education, Science and Culture.
N. K. also acknowledges the support by
Monbusho International Scientific Research Program.
K. V. acknowledges the support by the Japanese Government (Monbusho)
Scholarship Program as well.

\newpage
\begin{center}
APPENDIX  \par
\end{center}

Here we wish to prove the identity (12).
We first express  the products in (11)
by the Vandermonde determinants,
$$ \prod_{i=1}^r \frac{1}{x_{\alpha_i}-x_k}
= (-1)^r \frac{V^{(r)} (x_{\alpha_1} \ldots x_{\alpha_r})}
              {V^{(r+1)}(x_{\alpha_1} \ldots x_{\alpha_r} x_k)}
,
\eqno({\rm A}1) $$
$$ \prod_{j=1}^s \frac{1}{x_{\beta_j}-x_\ell}
= (-1)^s \frac{V^{(s)} (x_{\beta_1} \ldots x_{\beta_s})}
              {V^{(s+1)}(x_{\beta_1} \ldots x_{\beta_s} x_\ell)}
{}.
\eqno({\rm A}2)$$
Next, by using the binomial theorem we expand $(x_k-x_\ell)^{r+s-2}$.
After exchanging the order of finite sums,
we can rewrite the left-hand-side of Eq. (12) as follows
$$
\sum_{\{k \alpha_1 \ldots \alpha_r \} }
\sum_{\{\ell \beta_1 \ldots \beta_s \} }
b_q (k \alpha_1 \ldots \alpha_r \ell \beta_1 \ldots \beta_s)
$$
$$
=
\sum_{t=0}^{r+s-2}
\left ( \begin{array}{c}
r+s-2 \\ t \\ \end{array}
\right )  (-1)^{t+r}
\sum_{\{k \alpha_1 \ldots \alpha_r \} }
\frac{x_k^{r+s-t-2} V^{(r)} (x_{\alpha_1} \ldots x_{\alpha_r})}
        {V^{(r+1)}(x_{\alpha_1} \ldots x_{\alpha_r} x_k)}
\sum_{\{\ell \beta_1 \ldots \beta_s \} }
      \frac{x_\ell^t V^{(s)} (x_{\beta_1} \ldots x_{\beta_s})}
        {V^{(s+1)}(x_{\beta_1} \ldots x_{\beta_s} x_\ell)}
{}.
\eqno({\rm A}3)
$$
It is useful to introduce a following determinant
$$ W^{(r,u)} ( x_{\alpha_1} \ldots x_{\alpha_r} x_k )
= {\rm det} \left |
\begin{array}{ccccc}
1 & 1 & \ldots & 1 & 1 \\
x_{\alpha_1} & x_{\alpha_2} & \ldots & x_{\alpha_r} & x_k \\
\vdots & \vdots &  & \vdots & \vdots \\
x_{\alpha_1}^{r-1} & x_{\alpha_2}^{r-1} & \ldots &
x_{\alpha_r}^{r-1} & x_k^{r-1} \\
x_{\alpha_1}^u & x_{\alpha_2}^u & \ldots &
x_{\alpha_r}^u & x_k^u \\
\end{array}
  \right |  ,
\eqno({\rm A}4) $$
where $u=r+s-t-2$. The expansion of the determinant
$W^{(r,u)} ( x_{\alpha_1} \ldots x_{\alpha_r} x_k )$
with respect to the last row,
and the permutations of columns in the determinant
$V^{(r+1)}(x_{\alpha_1} \ldots x_{\alpha_r} x_k)$,
give the wanted sum over permutations
$k \leftrightarrow \alpha_i$, $i=1,2,\ldots,r$,
$$
\sum_{\{k \alpha_1 \ldots \alpha_r \} }
\frac{x_k^{r+s-t-2} V^{(r)} (x_{\alpha_1} \ldots x_{\alpha_r})}
     {V^{(r+1)}(x_{\alpha_1} \ldots x_{\alpha_r} x_k)}
=
\frac{W^{(r,r+s-t-2)} ( x_{\alpha_1} \ldots x_{\alpha_r} x_k )}
     {V^{(r+1)}(x_{\alpha_1} \ldots x_{\alpha_r} x_k)}
{}.
\eqno({\rm A}5)$$
Similarly, the sum over permutation
$\ell \leftrightarrow \beta_j$, $j=1,2,\ldots,s$, is
$$
\sum_{\{\ell \beta_1 \ldots \beta_s \} }
\frac{x_\ell^{t} V^{(s)} (x_{\beta_1} \ldots x_{\beta_s})}
     {V^{(s+1)}(x_{\beta_1} \ldots x_{\beta_s} x_\ell)}
=
\frac{W^{(s,t)} ( x_{\beta_1} \ldots x_{\beta_s} x_\ell )}
     {V^{(s+1)}(x_{\beta_1} \ldots x_{\beta_s} x_\ell)}
{}.
\eqno({\rm A}6) $$
If $ s-1 \leq t \leq r+s-2 $, then two rows of the determinant
$W^{(r,r+s-t-2)} ( x_{\alpha_1} \ldots x_{\alpha_r} x_k )$
are the same.
If $ 0 \leq t \leq s-1 $, then two rows of the determinant
$W^{(s,t)} ( x_{\beta_1} \ldots x_{\beta_s} x_\ell )$
are the same. Therefore, for any $t$ with
$ 0 \leq t \leq r+s-2 $ at least one of the sums over
permutations in Eq. (A3) is zero.
Consequently, the formula (12)
has been proven.

\newpage
\begin{center}
REFERENCES \par
\end{center}

\begin{description}

\item[{\rm [1]}]
F. Calogero,
J. Math. Phys. {\bf10}, 2197 (1969).

\item[{\rm [2]}]
B. Sutherland,
J. Math. Phys. {\bf12}, 246 (1971);
Phys. Rev.  A{\bf 4}, 2019 (1971);
{\it ibid.} A{\bf 5}, 1372 (1971).

\item[{\rm [3]}]
F. D. M. Haldane, Phys. Rev. Lett. {\bf 60}, 635 (1988);
{\it ibid.} {\bf 66}, 1529 (1991).

\item[{\rm [4]}]
B. S. Shastry,  Phys. Rev. Lett.  {\bf 60}, 639 (1988);
{\it ibid.} {\bf 69}, 164 (1992).

\item[{\rm [5]}]
Y. Kuramoto and H. Yokoyama, Phys. Rev. Lett. {\bf 67},
1338 (1991).

\item[{\rm [6]}]
Z. N. C. Ha and F. D. M. Haldane,
  Phys. Rev. B{\bf 46}, 9359 (1992).

\item[{\rm [7]}]
N. Kawakami, Phys. Rev. B{\bf 46}, 1005, 3192 (1992);
J. Phys. Soc. Jpn. {\bf 62}, 2270 (1993).

\item[{\rm [8]}]
A. P. Polychronakos, Phys. Rev. Lett. {\bf 69}, 703 (1992);
K. Hikami and M. Wadati, Phys. Lett. A{\bf 173}, 263 (1993).

\item[{\rm [9]}]
B. Sutherland and B. S. Shastry,
Phys. Rev. Lett. {\bf 71}, 5 (1993).

\item[{\rm [10]}]
Note that a multicomponent model
with a specific condition has been
discussed in M. Fowler and J. A. Minahan,
 Phys. Rev. Lett. {\bf 70}, 2325 (1993).

\item[{\rm [11]}]
J. H. F. Scott-Thomas, S. B. Field, M. A. Kastner, D. A. Antoniadis,
and H. I. Smith,
Phys. Rev. Lett. {\bf62}, 583 (1989).

\item[{\rm [12]}]
U. Meirav, M. A. Kastner, M. Heiblum, and S. J. Wind,
Phys. Rev. B{\bf40}, 5871 (1989).

\item[{\rm [13]}]
U. Meirav, M. A. Kastner, and S. J. Wind,
Phys. Rev. Lett. {\bf65}, 771 (1990).

\item[{\rm [14]}]
S. B. Field, M. A. Kastner, U. Meirav,
J. H. F. Scott-Thomas,
D. A. Antoniadis, H. I. Smith, and S. J. Wind,
Phys. Rev. B{\bf42}, 3523 (1990).

\item[{\rm [15]}]
H. van Houten and C. W. J. Beenakker,
Phys. Rev. Lett. {\bf63}, 1893 (1989);
C. W. J. Beenakker,
Phys. Rev. B{\bf44}, 1646 (1991).

\item[{\rm [16]}]
M. A. Kastner, S. B. Field, U. Meirav, J.H.F. Scott-Thomas,
D. A. Antoniadis, and H. I. Smith,
Phys. Rev. Lett. {\bf63}, 1894 (1989).

\item[{\rm [17]}]
Y. Meir, N. S. Wingreen, and P. A. Lee,
Phys. Rev. Lett. {\bf66}, 3048 (1991).

\item[{\rm [18]}]
S. Tewari,
Phys. Rev. B{\bf46}, 7782 (1992).

\item[{\rm [19]}]
N. F. Johnson and M. C. Payne,
Phys. Rev. Lett. {\bf70}, 1513 (1993);
{\it ibid.} {\bf70}, 3523 (1993).

\item[{\rm [20]}]
N. Kawakami, Kyoto preprint YITP/K-1013 (1993).

\item[{\rm [21]}]
A. A. Belavin, A. M. Polyakov and A. B. Zamolodchikov,
Nucl. Phys. {\bf B241}, 333 (1984).

\item[{\rm [22]}]
The reservoir is an ideal source and sink of electrons,
feeding the system up to the chemical potential $\mu$.
For details, see M. B\"{u}ttiker,
Phys. Rev. Lett. {\bf57}, 1761 (1986).

\end{description}

\end{document}